\documentstyle[12pt,epsfig,cite]{article}

\renewenvironment{thebibliography}[1]
 { \rm
   \begin{list}{\arabic{enumi}.}
    {\usecounter{enumi} \setlength{\parsep}{0pt}
     \setlength{\itemsep}{3pt} \settowidth{\labelwidth}{#1.}
     \sloppy
    }}{\end{list}}

\begin{document}

\baselineskip=16pt
\phantom{x}

\vglue 1.58 truein
\begin{flushleft}
{\bf NEW CLOCK COMPARISON TESTS OF LORENTZ SYMMETRY \\}
submitted to \emph{Proceedings of the 1999 Coral Gables Conference}
\end{flushleft}

\vglue 0.8cm
\begin{flushleft}
{\hskip 1 truein
Ronald L.\ Walsworth and David F.\ Phillips\\}
\bigskip
{\hskip 1 truein
Harvard-Smithsonian Center for Astrophysics\\}
{\hskip 1 truein
Cambridge, MA 02138\\}
{\hskip 1 truein
U.S.A.\\}
\end{flushleft}

\vglue 0.8cm

\noindent
{\bf INTRODUCTION}
\vglue 0.4 cm

Lorentz symmetry is a fundamental feature of modern descriptions of nature.
Lorentz transformations include both spatial rotations and boosts.
Therefore, experimental investigations of rotation symmetry provide
important tests of the framework of the standard model of particle physics
and single-metric theories of gravity \cite{will}. 

In particular, the minimal SU(3)$\times$SU(2)$\times$U(1) standard model successfully
describes particle phenomenology, but is believed to be the low energy limit
of a more fundamental theory that incorporates gravity.  While the
fundamental theory should remain invariant under Lorentz transformations,
spontaneous symmetry-breaking could result at the level of the standard
model in violations of local Lorentz invariance (LLI) and CPT (symmetry
under simultaneous application of Charge conjugation, Parity inversion, and
Time reversal) \cite{sachs}. 

\begin{figure}[tbp!]
\begin{center}
\begin{minipage}{.4\linewidth}  
\begin{center}
\epsfig{file=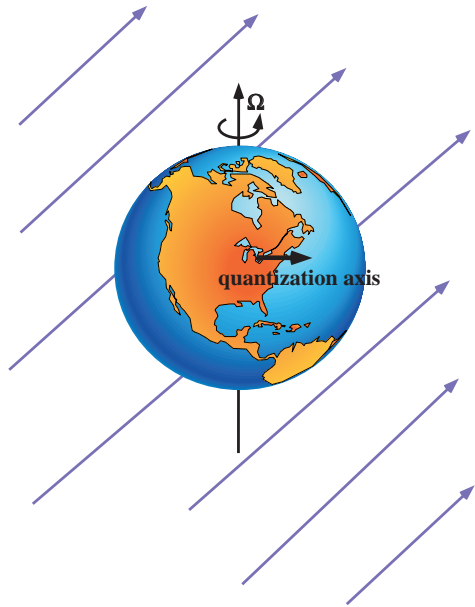,width=\textwidth,clip=,}
\end{center}
\end{minipage}
\hfill
\begin{minipage}{.4\linewidth}
\begin{center}
\caption{Bounds on LLI and CPT violation can be obtained by comparing 
  the frequencies of clocks as they rotate with respect to the fixed
  stars.  The standard model extension described in
  \cite{kl99a,ckpv,bkr9798,cfj,ck97jk99,bckp,bkr99,bkl00,bk00,cg}
  admits Lorentz-violating couplings of noble gas nuclei and hydrogen
  atoms to expectation values of tensor fields.  (Some of these
  couplings also violate CPT.) Each of the tensor fields may have an
  independent magnitude and orientation in space, to be limited by
  experiment.  The background arrows in this figure illustrate one
  such field.}
\label{f.LLIearth}
\end{center}
\end{minipage}
\end{center}
\end{figure}

Clock comparisons provide sensitive tests of rotation invariance and
hence Lorentz symmetry by bounding the frequency variation of a given
clock as its orientation changes, e.g., with respect to the fixed
stars \cite{kl99a}.  In practice, the most precise limits are obtained
by comparing the frequencies of two co-located clocks as they rotate
with the Earth (see Fig.\ \ref{f.LLIearth}).  Atomic clocks are
typically used, involving the electromagnetic signals emitted or
absorbed on hyperfine or Zeeman transitions.

We report preliminary results from two new atomic clock tests of LLI and CPT:

\renewcommand{\labelenumi}{(\arabic{enumi})}
\begin{enumerate}
\item Using a two-species ${}^{129}$Xe/${}^3$He Zeeman maser
\cite{chupp94,stoner96,bear98} we have placed a limit on CPT and LLI
violation of the neutron of nearly $10^{-31}$ GeV, improving by 
about a factor of three on the bound set by the most sensitive past 
clock comparison \cite{hl95,hl99}.  With further data-taking, 
we expect another factor of two improvement 
in sensitivity.
\item We have also employed atomic hydrogen masers to set an improved
  clean limit on LLI/CPT violation of the proton, at the level of
  nearly $10^{-27}$ GeV.
\end{enumerate}

\vglue 0.6 cm
\noindent
{\bf MOTIVATION}
\vglue 0.4 cm

Our atomic clock comparisons are motivated by a standard model
extension developed by Kosteleck\'y and others
\cite{kl99a,ckpv,bkr9798,cfj,ck97jk99,bckp,bkr99,bkl00,bk00,cg}.  This
theoretical framework accommodates possible spontaneous violation of
local Lorentz invariance (LLI) and CPT symmetry, which may occur in a
fundamental theory combining the standard model with gravity.  For
example, this might occur in string theory \cite{kskp}.  The standard
model extension is quite general: it emerges as the low-energy limit
of any underlying theory that generates the standard model and
contains spontaneous Lorentz symmetry violation \cite{ks89}.  The
extension retains the usual gauge structure and power-counting
renormalizability of the standard model.  It also has many other
desirable properties, including energy-momentum conservation, observer
Lorentz covariance, conventional quantization, and hermiticity.
Microcausality and energy positivity are expected.

This well-motivated theoretical framework suggests that small,
low-energy signals of LLI and CPT violation may be detectable in
high-precision experiments.  The dimensionless suppression factor for
such effects would likely be the ratio of the low-energy scale to the
Planck scale, perhaps combined with dimensionless coupling constants
\cite{kl99a,ckpv,bkr9798,cfj,ck97jk99,bckp,bkr99,bkl00,bk00,cg,kskp,ks89}.
A key feature of the standard model extension of Kosteleck\'y \emph{et
al.} is that it is at the level of the known elementary particles,
and thus enables quantitative comparison of a wide array of tests of
Lorentz symmetry.  In recent work the standard model extension has
been used to quantify bounds on LLI and CPT violation from
measurements of neutral meson oscillations \cite{ckpv}; tests of QED
in Penning traps \cite{bkr9798}; photon birefringence in the vacuum
\cite{cfj,ck97jk99}; baryogenesis \cite{bckp}; hydrogen and
antihydrogen spectroscopy \cite{bkr99}; experiments with muons
\cite{bkl00}; a spin-polarized torsion pendulum \cite{bk00};
observations with cosmic rays \cite{cg}; and atomic clock comparisons
\cite{kl99a}.  Recent experimental work motivated by this standard
model extension includes Penning trap tests by Gabrielse \emph{et al.} on
the antiproton and H$^-$ \cite{gab99}, and by Dehmelt \emph{et al.} on the
electron and positron \cite{deh99a,deh99b}, which place improved
limits on CPT and LLI violation in these systems.  Also, a re-analysis
by Adelberger, Gundlach, Heckel, and co-workers of existing data from
the ``E\"ot-Wash II'' spin-polarized torsion pendulum
\cite{adel99,harris98} sets the most stringent bound to date on CPT
and LLI violation of the electron: approximately $10^{-29}$ GeV
\cite{hec99}.

\vglue 0.6 cm
\noindent
{\bf ${}^{129}$Xe/${}^3$He MASER TEST OF CPT AND LORENTZ SYMMETRY}
\vglue 0.4 cm

\begin{figure}
\begin{center}
\epsfig{file=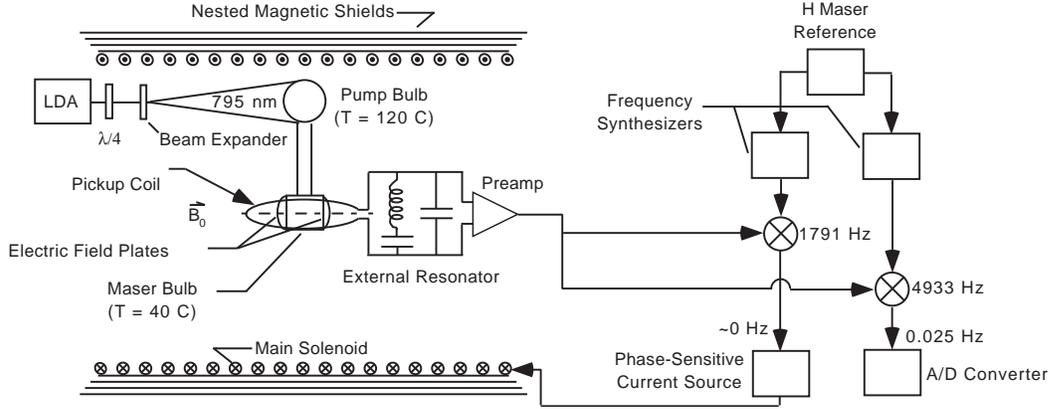,width=0.9\textwidth,clip=,}
\end{center}
\caption{Schematic of the ${}^{129}$Xe/${}^{3}$He Zeeman maser}
\label{f.DNGMschem}
\end{figure}

The design and operation of the two-species ${}^{129}$Xe/${}^{3}$He
maser has been discussed in recent publications
\cite{chupp94,stoner96,bear98}.  (See the schematic in Fig.\ 
\ref{f.DNGMschem}.) Two dense, co-habitating ensembles of ${}^{3}$He
and ${}^{129}$Xe atoms perform continuous and simultaneous maser
oscillations on their respective nuclear spin 1/2 Zeeman transitions
at approximately 4.9 kHz for ${}^{3}$He and 1.7 kHz for ${}^{129}$Xe
in a static magnetic field of 1.5 gauss.  This two-species maser
operation can be maintained indefinitely.  The population inversion
for both maser ensembles is created by spin exchange collisions
between the noble gas atoms and optically-pumped Rb vapor \cite{wh97}.
The ${}^{129}$Xe/${}^{3}$He maser has two chambers, one acting as the
spin exchange ``pump bulb'' and the other serving as the ``maser
bulb''.  This two chamber configuration permits the combination of
physical conditions necessary for a high flux of spin-polarized noble
gas atoms into the maser bulb, while also maintaining ${}^{3}$He and
${}^{129}$Xe maser oscillations with good frequency stability: ~ 100
nHz stability is typical for measurement intervals of  $\sim 1$ hour
\cite{bear98}.  (A single-bulb ${}^{129}$Xe/${}^{3}$He maser does not
provide good frequency stability because of the large Fermi contact
shift of the ${}^{129}$Xe Zeeman frequency caused by ${}^{129}$Xe-Rb
collisions \cite{rc98}.) Either of the noble gas species can serve as a
precision magnetometer to stabilize the system's static magnetic
field, while the other species is employed as a sensitive probe for
LLI- and CPT-violating interactions or other subtle physical
influences.  (For example, we are also using the
${}^{129}$Xe/${}^{3}$He maser to search for a permanent electric
dipole moment of ${}^{129}$Xe as a test of time reversal symmetry;
hence the electric field plates in Fig.\ \ref{f.DNGMschem}.)

We search for a signature of Lorentz violation by monitoring the
relative phases and Zeeman frequencies of the co-located ${}^{3}$He
and ${}^{129}$Xe masers as the laboratory reference frame rotates with
respect to the fixed stars.  We operate the system with the
quantization axis directed east-west in the earth's reference frame,
the ${}^{3}$He maser free-running, and the ${}^{129}$Xe maser
phase-locked to a signal derived from a hydrogen maser in order to
stabilize the magnetic field.  To leading order, the standard model
extension of Kosteleck\'y \emph{et al.} predicts that the Lorentz-violating
frequency shifts for the ${}^{3}$He and ${}^{129}$Xe maser are the
same size and sign \cite{kl99a}.  Hence the possible Lorentz-violating
effect would have the form:
\begin{equation}
\label{e.DNGMtwoShift}
\delta \nu_{\mathit{Lorentz}} = \delta \nu_{\mathit{He}}
\left[\gamma_{\mathit{He}} / \gamma_{\mathit{Xe}}  - 1 \right]^{-1},
\end{equation}
where $\delta \nu_{\mathit{Lorentz}}$ is the frequency shift
induced in the noble gas Zeeman transitions by the Lorentz-violating
interaction, $\delta \nu_{\mathit{He}}$ is the observed limit
to a frequency shift in the free-running ${}^3$He maser with the
period of a sidereal day ($\approx 23.93$ hours), and
$\gamma_{\mathit{He}} / \gamma_{\mathit{Xe}} \approx
2.75$ is the ratio of gyromagnetic ratios for ${}^3$He and
${}^{129}$Xe.

We acquired data for this experiment over a period of 30 days in
April, 1999 and 24 days in September, 1999.  (We are currently taking
additional data.)  We reversed the main magnetic field of the
apparatus every $\sim 4$ days to help distinguish possible
Lorentz-violating effects from diurnal systematic variations.  In
addition, we carefully assessed the effectiveness of the ${}^{129}$Xe
co-magnetometer, and found that it provides excellent isolation from
possible diurnally-varying ambient magnetic fields, which would not
average away with field reversals.  Furthermore, the relative phase
between the solar and sidereal day evolved almost $5\pi/6$ radians
between April and September; hence diurnal systematic effects from any
source would be reduced by averaging the results from the two existing
measurement sets.

\begin{figure}
\begin{center}
\epsfig{file=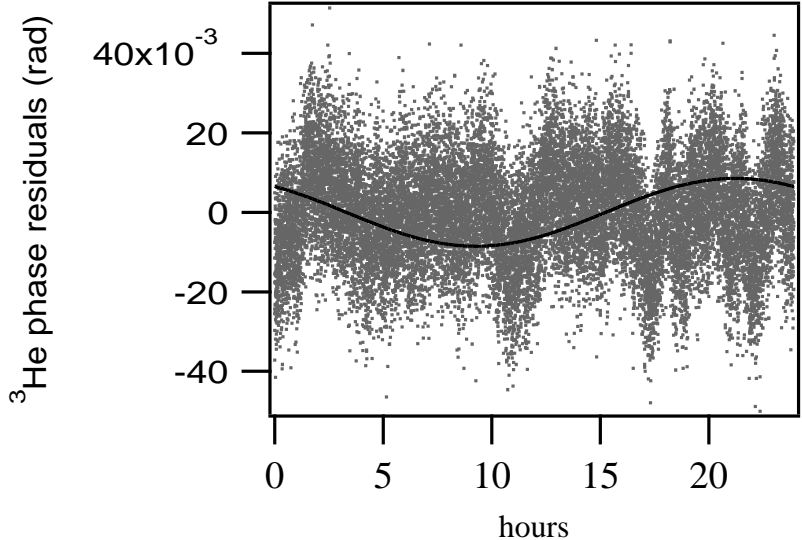,width=0.75\textwidth,clip=,}
\end{center}
\caption{Typical data from the LLI/CPT test using the 
${}^{129}$Xe/${}^{3}$He maser.  ${}^{3}$He maser
phase data residuals are shown for one sidereal day.  Larmor precession and
drift terms have been removed, and the best-fit sinusoid curve (with
sidereal-day-period) is displayed }
\label{f.DNGMdata}
\end{figure}

We analyzed each day's data in the April and September measurement
sets and determined the amplitude and phase of any sidereal-day-period
variation in the free-running ${}^{3}$He maser frequency.  (See Fig.\ 
\ref{f.DNGMdata} for an example of one day's data.) We employed a
linear least squares method to fit the free-running maser phase vs.\
time using a minimal model including: a constant (phase offset); a
linear term (Larmor precession); and cosine and sine terms with
sidereal day period.  For each day's data, we included terms
corresponding to quadratic and maser amplitude-induced phase drift if
they significantly improved the reduced $\chi^2$ \cite{bev}.  As a
final check, we added a \emph{faux} Lorentz-violating effect of known phase
and amplitude to the raw data and performed the analysis as before.
We considered our data reduction for a given sidereal day to be
successful if the synthetic physics was recovered and there was no
significant change in the covariance matrix generated by the fitting
routine.

Using the April and September measurement sets, we found no
statistically significant sidereal variation of the free-running
${}^{3}$He maser frequency.  At the one-sigma confidence level, we
determined a bound of approximately 80 nHz on
$\delta\nu_{\mathit{Lorentz}}$, the magnitude of a possible LLI- and
CPT-violating effect.  Kosteleck\'y and Lane report that the nuclear
Zeeman transitions of ${}^{129}$Xe and ${}^{3}$He are primarily
sensitive to Lorentz-violating couplings of the neutron, assuming the
correctness of the Schmidt model of the nuclei \cite{kl99a}.  Thus our
search for a sidereal-period frequency shift of the free-running
${}^{3}$He maser ($\delta\nu_{\mathit{He}}$) provides a bound to the
following parameters characterizing the magnitude of LLI/CPT
violations in the standard model extension:

\begin{eqnarray}
\label{e.DNGMkosShift}
\left|-3.5 \tilde{b}^n_J + 0.012 \tilde{d}^n_J
+ 0.012 \tilde{g}^n_{D,J} \right| \le 2\pi \delta \nu_{\mathit{He,J}} & & 
({}^{129}\mathrm{Xe}/{}^{3}\mbox{He maser})
\end{eqnarray}
Here $J = X,Y$ denotes spatial indices in a non-rotating frame, with
$X$ and $Y$ oriented in a plane perpendicular to the Earth's rotation
axis and we have taken $\hbar = c = 1$.  The parameters
$\tilde{b}^n_J$, $\tilde{d}^n_J$, and $\tilde{g}^n_{D,J}$ describe the
strength of Lorentz-violating couplings of the neutron to possible
background tensor fields.  $\tilde{b}^n_J$ and $\tilde{g}^n_{D,J}$
correspond to couplings that violate both CPT and LLI, while
$\tilde{d}^n_J$ corresponds to a coupling that violates LLI but not
CPT.  All three of these parameters are different linear combinations
of fundamental parameters in the underlying relativistic
Lagrangian of the standard model extension
\cite{kl99a,ckpv,bkr9798,cfj,ck97jk99,bckp,bkr99,bkl00,bk00}.

It is clear from Eqn.\ (\ref{e.DNGMkosShift}) that the
${}^{129}$Xe/${}^{3}$He clock comparison is primarily sensitive to
LLI/CPT violations associated with the neutron parameter
$\tilde{b}^n_J$.  Similarly, the most precise previous search for
LLI/CPT violations of the neutron, the ${}^{199}$Hg/${}^{133}$Cs
experiment of Lamoreaux, Hunter \emph{et al.} \cite{hl95,hl99}, also
had principal sensitivity to $\tilde{b}^n_J$ at the following level
\cite{kl99a}:
\begin{eqnarray}
\label{e.CsKosShift}
\left| \frac{2}{3} \tilde{b}^n_J +\left\{ \mbox{small terms} \right\} 
\right| \le 2 \pi \delta \nu_{\mathit{Hg,J}}  & & 
({}^{199}\mathrm{Hg}/{}^{133}\mathrm{Cs}).
\end{eqnarray}
In this case, the experimental limit, $\delta \nu_{\mathit{Hg,J}}$, was a
bound of 55 nHz (one-sigma) on a sidereal-period variation of the
${}^{199}$Hg nuclear Zeeman frequency, with the ${}^{133}$Cs electronic
Zeeman frequency serving as a co-magnetometer.

Therefore, in the context of the standard model extension of
Kosteleck\'y and co-workers \cite{kl99a}, our ${}^{129}$Xe/${}^{3}$He
maser measurement improves the constraint on $\tilde{b}^n_J$ to nearly
$10^{-31}$ GeV, or about three times better than the
${}^{199}$Hg/${}^{133}$Cs clock comparison \cite{hl95,hl99}.  Note
that the ratio of this limit to the neutron mass ($10^{-31}
\mathrm{GeV}/m_n \sim 10^{-31}$) compares favorably to the
dimensionless suppression factor $m_n / M_{\mathit{Planck}} \sim
10^{-19}$ that might be expected to govern spontaneous symmetry
breaking of LLI and CPT originating at the Planck scale.  We expect a
factor of two improvement in the sensitivity to LLI/CPT-violation with
the completion of ${}^{129}$Xe/${}^{3}$He maser data taking in spring,
2000.  We also expect more than an order of magnitude improvement in
sensitivity using a new device currently under development in our
laboratory: the ${}^{21}$Ne/${}^3$He Zeeman maser.

\vglue 0.6 cm
\noindent
{\bf HYDROGEN MASER TEST OF CPT AND LORENTZ SYMMETRY}
\vglue 0.4 cm

\begin{figure}
\begin{center}
\epsfig{file=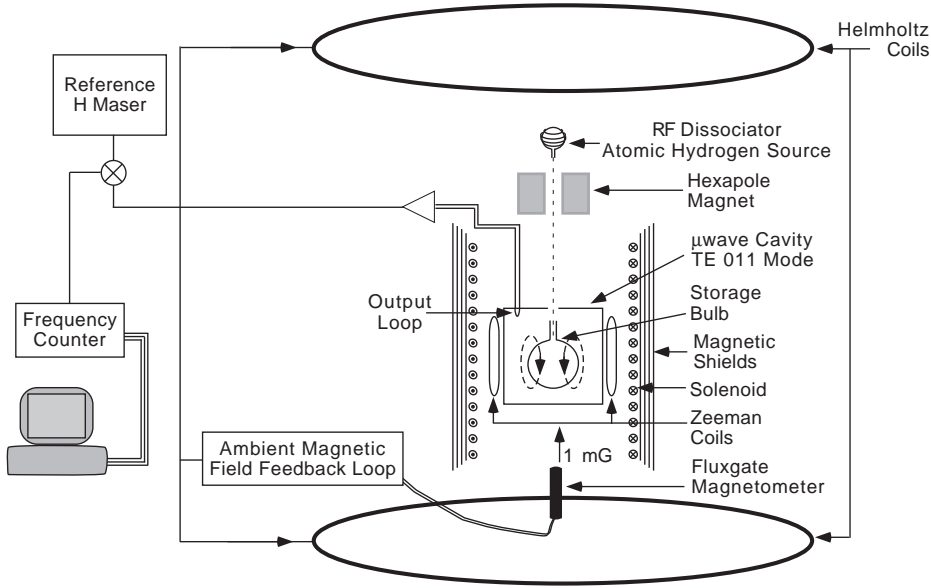,width=0.8\textwidth,clip=,}
\end{center}
\caption{Schematic of the H maser in its ambient field stabilization loop.}
\label{f.Hschem}
\end{figure}

The hydrogen maser is an established tool in precision tests of
fundamental physics \cite{happs}. Hydrogen masers operate on the
$\Delta F = 1$, $\Delta m_F = 0$ hyperfine transition in the ground
state of atomic hydrogen \cite{kgrvv}.  Hydrogen molecules are
dissociated into atoms in an RF discharge, and the atoms are state
selected via a hexapole magnet (Fig.\ \ref{f.Hschem}). The high field
seeking states, ($F=1$, $m_F = +1,0$) are focused into a Teflon coated
cell which resides in a microwave cavity resonant with the $\Delta
F=1$ transition at 1420 MHz. The $F=1$, $m_F=0$ atoms are stimulated
to make a transition to the $F=0$ state by the field of the cavity.  A
static magnetic field of $\sim 1$ milligauss is applied to maintain
the quantization axis of the H atoms.

\begin{figure}
\begin{center}
\epsfig{file=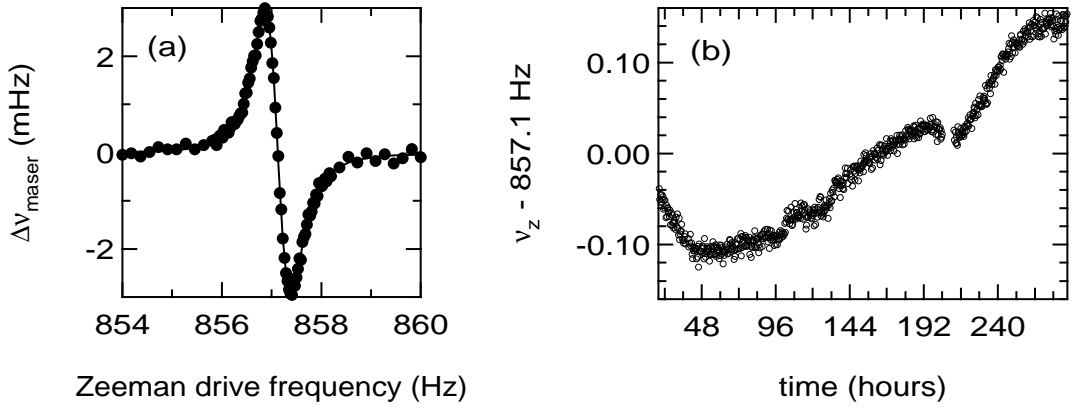,width=\textwidth,clip=,}
\end{center}
\caption{ 
  (a) An example of a double resonance measurement of the $F=1$,
  $\Delta m_F = \pm 1$ Zeeman frequency in the hydrogen maser. The
  change from the unperturbed maser frequency is plotted versus the
  driving field frequency.  (b) Zeeman data from 11 days of the
  LLI/CPT test using the H maser.  }
\label{f.Hdata}
\end{figure}

The hydrogen transitions most sensitive to potential CPT and LLI
violations are the $F=1$, $\Delta m_F= \pm 1$ Zeeman transitions.  In
the $0.6$ mG static field applied for these measurements, the Zeeman
frequency is $\nu_Z \approx 850$ Hz.  We utilize a double resonance
technique to measure this frequency with a precision of $\sim 1$ mHz
\cite{andresen}. We apply a weak magnetic field perpendicular to the
static field and oscillating at a frequency close to the Zeeman
transition. This audio-frequency driving field couples the three
sublevels of the $F=1$ manifold of the H atoms.  Provided a population
difference exists between the $m_F= \pm 1$ states, the energy of the
$m_F=0$ state is altered by this coupling, thus shifting the measured
maser frequency in a carefully analyzed manner \cite{andresen}
described by a dispersive shape (Fig.\ \ref{f.Hdata}(a)). Importantly,
the maser frequency is unchanged when the driving field is exactly
equal to the Zeeman frequency. Therefore, we determine the Zeeman
frequency by measuring the driving field frequency at which the maser
frequency in the presence of the driving field is equal to the
unperturbed maser frequency.

The $F=1$, $\Delta m_F=\pm 1$ Zeeman frequency is directly
proportional to the static magnetic field, in the small-field limit.
Four layers of high permeability ($\mu$-metal) magnetic shields
surround the maser (Fig.\ \ref{f.Hschem}), screening external field
fluctuations by a factor of $32 \, 000$. Nevertheless, external
magnetic field fluctuations cause remnant variations in the observed
Zeeman frequency. As low frequency magnetic noise in the neighborhood
of this experiment is much larger during the day than late at night,
the measured Zeeman frequency could be preferentially shifted by this
noise (at levels up to $\sim 0.5$ Hz) with a 24 hour periodicity which
is difficult to distinguish from a true sidereal signal in our
relatively short data sample.  Therefore, we employ an active
stabilization system to cancel such magnetic field fluctuations (Fig.\ 
\ref{f.Hschem}). A fluxgate magnetometer placed as close to the maser
cavity as possible controls large (8 ft.\ dia.) Helmholtz coils
surrounding the maser via a feedback loop to maintain a constant
ambient field.  This feedback loop reduces the fluctuations at the
sidereal frequency to below the equivalent of 1 $\mu$Hz on the Zeeman
frequency at the location of the magnetometer.

The Zeeman frequency of a hydrogen maser was measured for 11 days in
November, 1999. During this period, the maser remained in a closed,
temperature controlled room to reduce potential systematics from
thermal drifts which might be expected to have 24 hour periodicities.
The feedback system also maintained a constant ambient magnetic field.
Each Zeeman measurement took approximately 20 minutes to acquire and
was subsequently fit to extract a Zeeman frequency (Fig.\ 
\ref{f.Hdata}(a)). Also monitored were maser amplitude, residual
magnetic field fluctuation, ambient temperature, and current through
the solenoidal coil which determines the Zeeman frequency (Fig.\ 
\ref{f.Hschem}).

The data for all 11 days (Fig.\ \ref{f.Hdata}(b)) were then fit to
extract the sidereal-period sinusoidal variation of the Zeeman
frequency. In addition to the sinusoid, piecewise linear terms (whose
slopes were allowed to vary independently for each day) were used to
model the slow remnant drift of the Zeeman frequency (roughly 250 mHz
over 11 days). A one sigma limit was established for a sidereal-period
variation of the hydrogen $F=1$, $\Delta m_F = \pm 1$ Zeeman
frequency: $\delta \nu_Z^H \le 0.7$ mHz. Expressed in terms of energy,
this is a shift in the Zeeman splitting of less than $3 \cdot
10^{-27}$ GeV.


%
%


The hydrogen atom is directly sensitive to LLI and CPT violations of
the proton and the electron. Following the notation of reference
\cite{bkr99}, one finds that a limit on a sidereal-period modulation
of the Zeeman frequency ($\delta \nu_Z^H$) provides a bound to the
following parameters characterizing the magnitude of LLI/CPT
violations in the standard model extension of Kosteleck\'y and
co-workers:
\begin{equation}
\left| b^e_3 + b^p_3 - d^e_{30} m_e - d^p_{30} m_p
- H^e_{12} - H^p_{12} \right| \le 2 \pi \delta \nu_Z^H 
\label{e.BKRshift}
\end{equation}
for the low static magnetic fields at which we operate. (Again, we
have taken $\hbar=c=1$.) The terms $b^e$ and $b^p$ describe the
strength of background tensor field couplings that violate CPT and LLI
while the $H$ and $d$ terms describe couplings that violate LLI but
not CPT \cite{bkr99}. The subscript 3 in Eqn.\ (\ref{e.BKRshift})
indicates the direction along the quantization axis of the apparatus,
which is vertical in the lab frame but rotates with respect to the
fixed stars with the period of the sidereal day.

As in refs.\ \cite{kl99a,deh99a}, we can
re-express the time varying change in the hydrogen Zeeman frequency in terms of
parameters expressed in a non-rotating frame as
\begin{equation}
2 \pi \delta \nu_{Z,J}^H = \left( \tilde{b}^p_J + \tilde{b}^e_J \right)
\sin{\chi}.
\label{e.Hnonrot}
\end{equation}
where $\tilde{b}^w_J = b^w_j -d^w_{j0} m_w - \frac{1}{2}
\epsilon_{JKL} H^w_{KL}$, $J=X,Y$ refers to non-rotating spatial
indices in the plane perpendicular to the rotation vector of the
earth, $w$ refers to either the proton or electron parameters, and
$\chi=42^\circ$ is the latitude of the experiment.

As noted above, a re-analysis by Adelberger, Gundlach, Heckel, and
co-workers of existing data from the ``E\"ot-Wash II'' spin-polarized
torsion pendulum \cite{adel99,harris98} sets the most stringent bound
to date on CPT and LLI violation of the electron: $\tilde{b}^e_J \le
10^{-29}$ GeV \cite{hec99}.  Therefore, in the context of the standard
model extension of Kosteleck\'y and co-workers \cite{bkr99,kl99a} the
H maser measurement to date constrains LLI and CPT violations of the
proton parameter $\tilde{b}^p_J \le 4 \cdot 10^{-27}$ GeV at the one
sigma level. This preliminary limit is comparable to that derived from
the ${}^{199}$Hg/${}^{133}$Cs experiment of Lamoreaux, Hunter \emph{et
  al.} \cite{hl95,hl99} but in a much cleaner system (the hydrogen
atom nucleus, a proton, as opposed to the ${}^{199}$Hg and
${}^{133}$Cs nuclei).  The experiment is in progress with more data
taking expected in early spring, 2000. Further studies of systematics will
be performed including reversal of the quantization axis as well as a
change in the magnitude of the applied field.

\vglue 0.6 cm
\noindent
{\bf CONCLUSIONS}
\vglue 0.4 cm

Precision comparisons of atomic clocks provide sensitive tests of
Lorentz and CPT symmetries, thereby probing extensions to the standard
model \cite{kl99a,ckpv,bkr9798,cfj,ck97jk99,bckp,bkr99,bkl00,bk00,cg}
in which these symmetries can be spontaneously broken.  Measurements
using the two-species ${}^{129}$Xe/${}^3$He Zeeman maser constrain
violations of CPT and Lorentz symmetry of the neutron at nearly the
$10^{-31}$ GeV level. Measurements with atomic hydrogen masers provide
clean tests of CPT and Lorentz symmetry violation of the proton at
nearly the $10^{-27}$ GeV level. Improvements in both experiments are
being pursued.

\vglue 0.6 cm
\noindent
{\bf ACKNOWLEDGMENTS}
\vglue 0.4 cm

Richard Stoner, David Bear, Marc Humphrey, Edward Mattison, and Robert
Vessot are collaborators on the research described in this report.  We
gratefully acknowledge the encouragement and active support of these
projects by Alan Kosteleck\'y, and technical assistance by Marc
Rosenberry and Timothy Chupp.  Development of the
${}^{129}$Xe/${}^3$He Zeeman maser was supported by a NIST Precision
Measurement Grant. Support for the Lorentz violation tests was
provided by NASA grant NAG8-1434, ONR grant N00014-99-1-0501, and the
Smithsonian Institution Scholarly Studies Program.


\vglue 0.6 cm
\noindent
{\bf REFERENCES}
\vglue 0.4 cm

\end{document}